\documentclass[prb,superscriptaddress,floatfix,twocolumn,showpacs,amsfonts,amsmath,amssymb]{revtex4}
\usepackage{graphicx} 
\usepackage{dcolumn} 
\usepackage{bm} 
\usepackage{color}

\begin{document}

\title{The origin of hour-glass magnetic dispersion in underdoped
cuprate superconductors}

\author{Y. A. Kharkov}
\affiliation{School of Physics, University of New South Wales, Sydney 2052, Australia}
\author{O. P. Sushkov}
\affiliation{School of Physics, University of New South Wales, Sydney 2052, Australia}

\date{\today}

\begin{abstract}
  In the present work we explain the hour-glass magnetic dispersion in
  underdoped cuprates. The dispersion arises due to the Lifshitz-type magnetic
  criticality.  Superconductivity also plays a role, but the role is secondary.
  We list six major experimental observations related to the hour-glass
  and  explain all of them.
  The theory provides a unified picture of the evolution of
  magnetic excitations in various cuprate families, including ``hour-glass''
  and ``wine-glass'' dispersions and an emergent static incommensurate order.
  We propose the Lifshitz spin liquid ``fingerprint'' sum rule, and show that the latest data confirm the validity of the sum rule.
\end{abstract}


\maketitle

\section{Introduction}
\begin{figure}[h!]
  \includegraphics[scale=0.17]{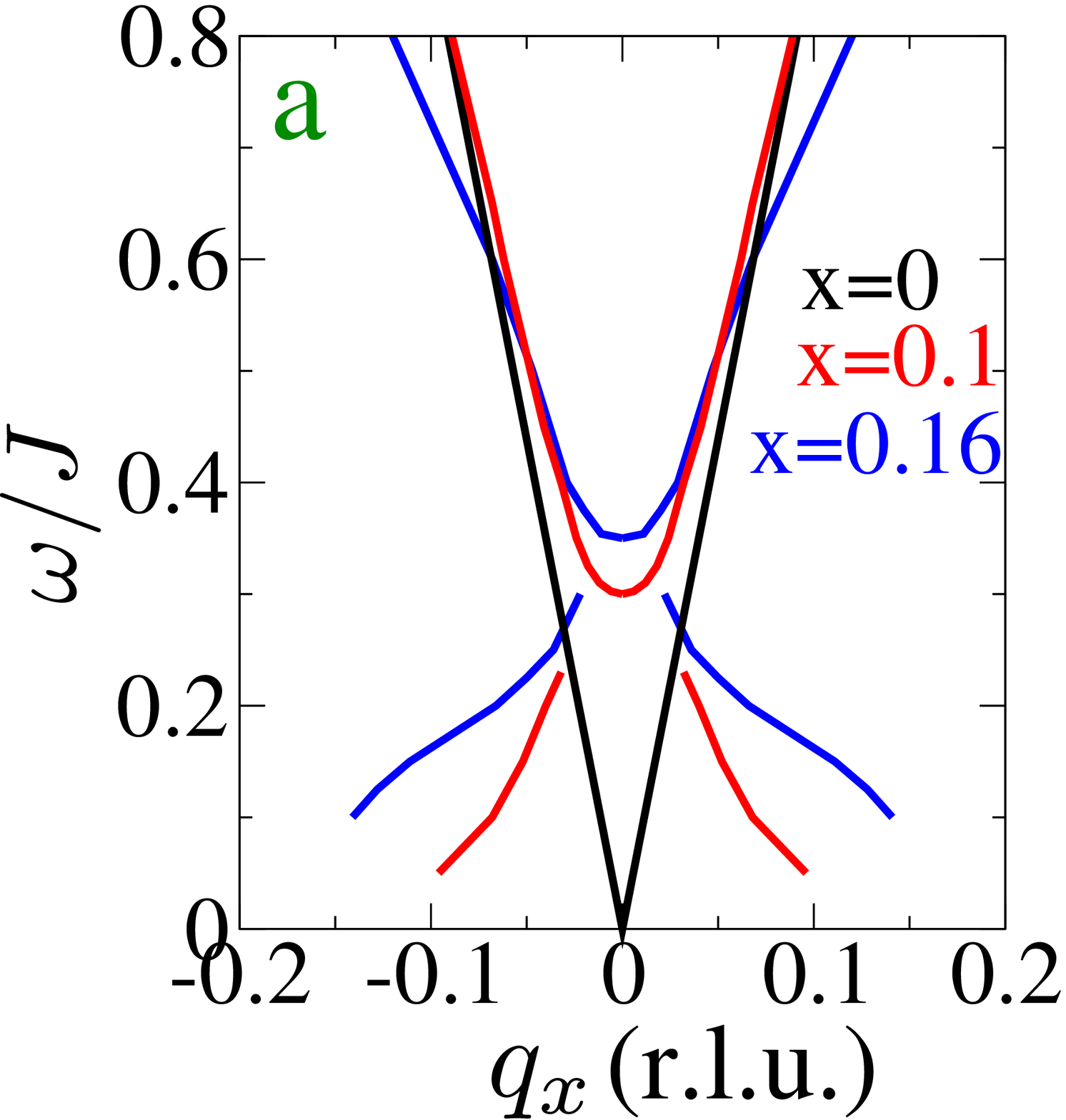}  
  \hspace{10pt}
  \includegraphics[scale=0.2]{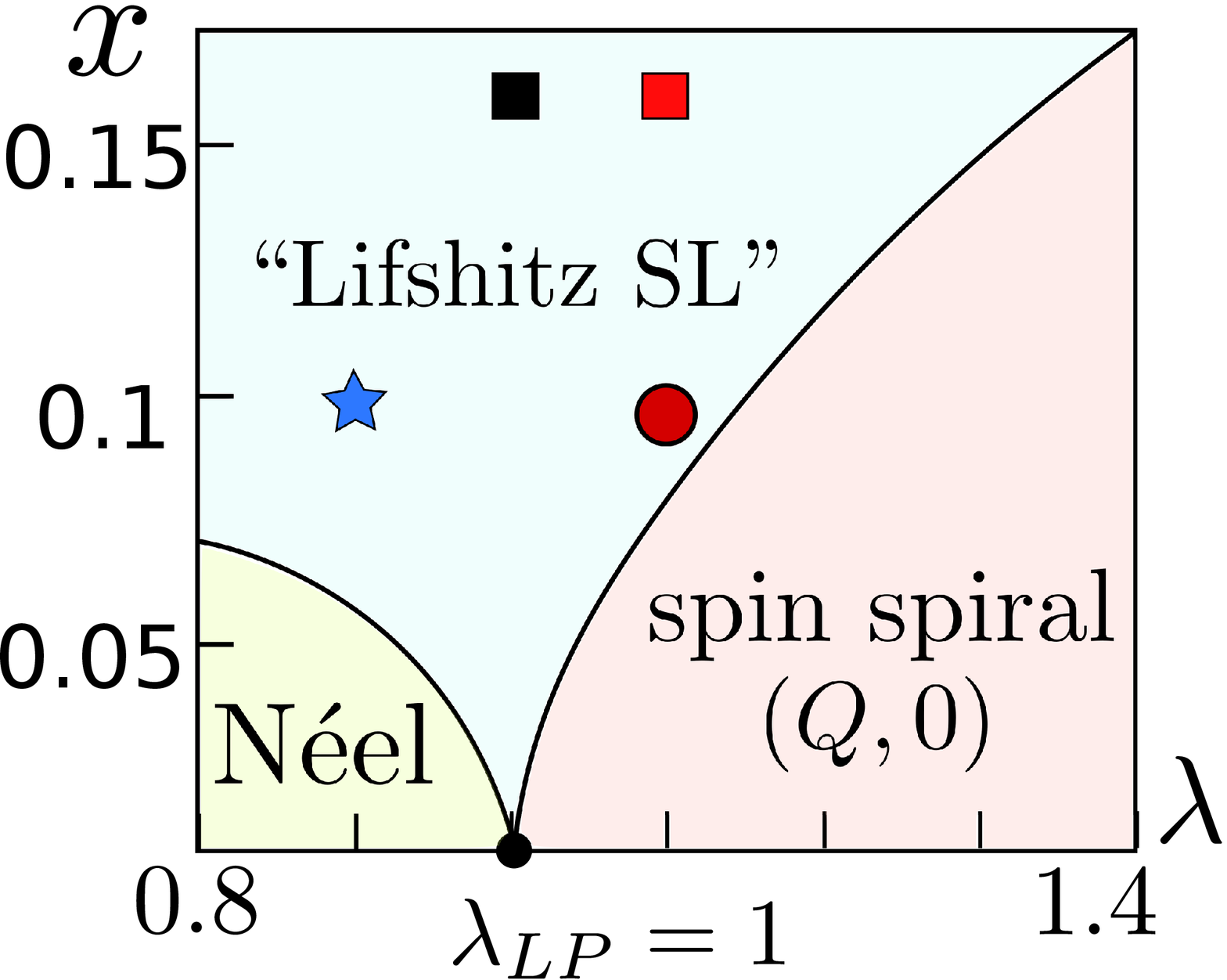}  
  \caption{Panel (a): Theoretical HG dispersions for $\lambda=1.1$ and
    for two values of doping,
    $x=0$.1 (red) and $x=0$.16 (blue). The black  line represents the
    magnon dispersion
    in the parent undoped compound ($x=0$).
Panel (b): Zero temperature $\lambda-x$ phase diagram of extended $t-J$ model consists of
    three phases, Neel, Lifshitz spin liquid, Spin  Spiral~\cite{Kharkov2018a}.
    The tricritical Lifshiz point is $\lambda=1$, $x=0$.
    The squares, the circle, and the star
    are the points considered in the text as examples.
    }
\label{HGf}
\end{figure}

The ``hour-glass'' (HG) dispersion, observed in inelastic neutron scattering,
is a generic property of hole doped high temperature cuprate
superconductors~\cite{Arai1999,Bourges2000,Hayden2004,Tranquada2004,Hinkov2007},
for a review see Ref.~\cite{Fujita2012}
The dispersion shown in Fig.\ref{HGf}a consists of the upper and
the lower branches, the  so called $(\pi,\pi)$ ``spin resonance'' separates
the two branches. 
In this work we shift momentum origin to $(\pi,\pi)$,
see Fig.\ref{fig:PD}a, so our $q=0$ corresponds to $(\pi,\pi)$ in neutron
scattering. This shift is convenient for theory and quite often is
used  in neutron scattering papers.
The HG dispersion is a major effect of strong electron correlations.
While there is a general feeling that the upper part of the HG is
due to localised spins and the lower part is related to
itinerant holes~\cite{Vojta2006,Fujita2012},
there is no understanding of the mechanism of this phenomenon in spite of two
decades of efforts.
There is a set of observations that must be explained, here they are: \\

\noindent
{\it 
{\bf (O1)}   The lower part of the HG shrinks to zero when doping is
decreasing, $x\to 0$, Ref.~\cite{Fujita2012}\\

\noindent
{\bf (O2)} In optimally doped cuprates, $x \approx 0.15$, the lower part of
the HG is observed
in the superconducting (SC) state and disappears in the normal (N) state,
Ref.\cite{Keimer1995}\\

\noindent
{\bf (O3)} Contrary to {\bf (O2)}, in underdoped cuprates
the lower part of HG is almost the same or exactly the same
in the SC state and the N state just above $T_c$,
see Refs.\cite{Keimer1997,Keimer1997a,Chan2016,Fujita2012}.
Moreover, the HG and the $(\pi,\pi)$ resonance were recently
observed~\cite{Keren2017} in the insulating
La$_{2-x}$Sr$_{x}$CuO$_4$ at $x=0.0192$ where SC does not exists.\\

\noindent
{\bf (O4)} The upper part of HG is always almost the same in the SC state
 and in the N state, and the slope of the upper part decreases
with doping, Fig.\ref{HGf}a.\\

\noindent
{\bf (O5)} In heavily underdoped cuprates the lower part of HG propagates
down to zero energy   resulting in an emergent static incommensurate
magnetic order~\cite{Yamada98,Haug2010}.\\

\noindent
{\bf (O6)}  All cuprate families are microscopically similar,
values of the superexchange and hopping matrix elements are close.
At the same time details of the lower part of the HG dispersion varies across different cuprate families. Moreover,
in underdoped HgBa$_2$CuO$_{4+\delta}$ the HG evolves to the
``wine-glass''\cite{Chan2016,Chan16a}.
}

Theoretical models of the magnetic dispersion and the $(\pi,\pi)$
resonance are split into two classes, models based on the normal
Fermi liquid picture with a large Fermi surface and usual electrons with
spin~\cite{Abanov2002,Onufrieva2002,Sherman2003,Eremin2005,Eremin2012},
and models based on the picture of a doped Mott insulator with a small Fermi
surface and spinless holons~\cite{Milstein2008,James2012}.
All early models have been motivated by experiment\cite{Keimer1995}
in optimally doped YBa$_2$Cu$_3$O$_7$ and belong to the
first class. In this approach the $(\pi,\pi)$ resonance is explained as a
spin exciton  in the d-wave SC phase.
These models are consistent with observation {\bf (O2)}, but
inconsistent with all other observations which appeared later and that
indicate that SC is not essential.
In light of this inconsistency the spin exciton model was modified by
artificial introduction of localised spins in the normal Fermi liquid
model~\cite{Sherman2003,Eremin2012}.
This modification partially explains the observation {\bf (O4)} (in addition
to {\bf (O2)}), but is still inconsistent with all other
observations.

The second theoretical approach based on the picture of a doped Mott
insulator was developed later after the
low doping data were obtained.
This approach naturally explains the observation {\bf (O1)}.
The model of Ref.\cite{Milstein2008}
is based on the picture of static spin spiral~\cite{Shraiman90}.
This model explains the observations {\bf (O1),(O3),(O5)}, but fails
in all other points.
The model of Ref.\cite{James2012} explains {\bf (O1)} and partially  
{\bf (O4)}, but fails in all other points.
Thus, the theoretical situation is unsatisfactory.

In the present work we pursue the approach of a lightly doped Mott insulator.
There are four major experimental facts  supporting the Mott insulator approach,
here they are.\\

\noindent
    {\bf (F1)} {\it According to NMR the nearest site
    antiferromagnetic exchange, $J \approx 125$meV, is doping
    independent~\cite{Imai1993}.}\\
    
\noindent
    {\bf (F2)} {\it The second fact  is the observation {\bf (O1)} from the HG
      list presented above. It is hardly possible to shrink the HG  to zero
      at zero doping in any model but doped Mott insulator.}\\

\noindent
    {\bf (F3)} {\it RIXS data indicate that the high energy
      magnons, $\omega \sim 200-300meV$,  in doped compounds are practically
      the same as
in undoped ones,
this includes both the dispersion and the spectral weight~\cite{LeTacon11}.
}\\

\noindent
{\bf (F4)} {\it The momentum integrated structure factor
$S(\omega)$ measured in neutron scattering at $\omega \approx 50-80$meV 
in doped compounds is practically the same as in undoped ones. We will demonstrate this
observation at the end of this paper.}\\

\noindent
These four facts unambiguously favour the Mott insulator approach.

The Mott insulator approach  necessarily implies a small Fermi surface,
Fig.\ref{fig:PD}a,
and this immediately leads to two conclusions that are evident without
calculations. The {\bf first} conclusion concerns superconductivity.
The Fermi energy is proportional to the doping x, $\epsilon_F \sim xJ$. 
For optimal doping, $x=0.15$, the Fermi energy is $\epsilon_F \approx 35$meV.
On the other hand we know experimentally that the superconducting gap
is  $\Delta_{SC} \approx 30$meV. Thus, all cuprates are in the strong
coupling limit, $\epsilon_F \approx \Delta_{SC}$.
The {\bf second} conclusion concerns the spin liquid ground state.
Consistently with small Fermi surface the number of charge carriers measured
via Hall effect is equal to the doping $x \ll 1$.
The number of uncompensated spins in a doped Mott insulator is $1-x$, so
unlike a normal metal the number of spins is much
larger than the number of charge carriers. We also know that the static
magnetic order disappears above several per cent doping,
when $x \ll 1-x$.
These points indicate that spin and charge are separated and
that quantum spin fluctuations 'melt' the static magnetic order to a spin liquid
(SL).
Of course the notion of SL in cuprates is not new, the same motivation is
behind the RVB SL model~\cite{Anderson}.

The present work is based on the recent progress  in understanding
of the SL state of cuprates~\cite{Kharkov2018a}.
The  SL in cuprates is different to the RVB model. It is the quantum
critical Ioffe-Larkin type SL (``Lifshitz SL'')~\cite{Kharkov2018a}.
This insight allows us to perform calculations and to
explain all properties of the HG.
There are the following sections in the paper.
II. Magnetic response in the spin liquid phase.
III. Calculated q-scans of the spectral function at optimal doping.
IV. Magnetic criticality mechanism of the hour-glass dispersion.
V. Is there a hole in the hour-glass?
VI. Calculated q-scans of the spectral function in the underdoped case.
Emergent incommensurate magnetic order.
VII Wine-glass dispersion.
VIII. Lifshitz spin liquid fingerprint relation.
IX. Conclusions.
Technical details are presented in Appendices A,B,C.

\section{Magnetic response in the spin liquid phase}
We start with the zero temperature $\lambda-x$ phase diagram from
Ref.~\cite{Kharkov2018a} presented in Fig.\ref{HGf}b.
The dimensionless parameter $\lambda$ defined as
\begin{eqnarray}
\label{ll}
\lambda=\frac{2g^2m^*}{\pi\rho_s} 
\end{eqnarray}
plays a crucial role in the theory, it controls  magnetic criticality.
Here $m^*$ is the holon effective mass,
$\rho_s$ is bare spin stiffness, and $g$ is the holon-magnon coupling constant,
for details see Appendix A.
The Lifshitz quantum tricritical point is $\lambda=1$, $x=0$.
Three phases meet at the tricritical point, the collinear N\'eel phase,
Lifshitz SL, the spin spiral state,
see Appendix B.
The Lifshitz SL is characterised by a parameter $\Delta$
that we term the SL gap. It is worth noting that in the exact sense
the SL is gapless. Below we also introduce and explain the spin pseudogap
$\Delta_s$.
Besides the Lifshitz quantum tricritical point
there is another critical point, $\lambda=2$, related to the phase
separation, see Appendix A.
Our analysis  is based on the extended $t-J$ model
which is described  by the antiferromagnetic exchange  $J\approx 125$meV
and hopping parameters.
The relation between parameters of the extended $t-J$ model and $\lambda$ is discussed in Appendix A.
Having the nearest site hopping parameter $t$ fixed, $t/J\approx 3$, one can
vary distant hopping parameters.
In principle this results in variation of $\lambda$ in a very broad range,
$0 < \lambda < \infty$.
Using values of the hopping matrix elements obtained in LDA
calculations~\cite{andersen95} one can obtain the following range for the criticality parameter
$\sim 0.7 \leq \lambda \leq 2$, see Appendix A. Nevertheless this range of
$\lambda$ is too wide,
it is even sufficient to drive the system to the phase separation.
Experiments indicate that most cuprates are magnetically disordered
except of the emergent magnetism at very low doping.
This observation allows us to restrict the range to 
approximately $\lambda=1.1\pm 0.3$ shown in Fig.\ref{HGf}b.
The criticality parameter $\lambda$ can vary  from one cuprate family to another and can slightly depend on doping.

To avoid misunderstanding we note that  the magnetic criticality we are
talking about is unrelated to the quantum critical point at doping
$x=x^*\approx 0.2$ (the endpoint of the ``pseudogap'' regime)
where presumably the small Fermi surface is transformed
to the large one~\cite{Taillefer2010}. We only consider $x < x^*$ and
claim that all hole doped cuprates are close to the Lifshitz magnetic
criticality. One can also call it the ``hidden'' criticality.
It is hidden in the sense that unlike doping, the parameter $\lambda$ cannot
be directly measured.

The first message of our paper is that the HG dispersion is a direct
consequence of the SL gap and the Lifshitz magnetic criticality.
 SC plays a secondary role, it only influences the particle-hole
decay phase space and narrows the spectral lines in the lower part of HG.
The first message resolves the generic problems {\bf (O1)-(O5)} listed in the
introduction.
The second message is that due to proximity to the quantum critical point
a small ($\sim 10\%$) variation of $\lambda$ results in  sizeable
change of the lower part of HG. This explains the point {\bf (O6)}
from the observation list.

\begin{figure}[h!]
  \includegraphics[scale=0.19]{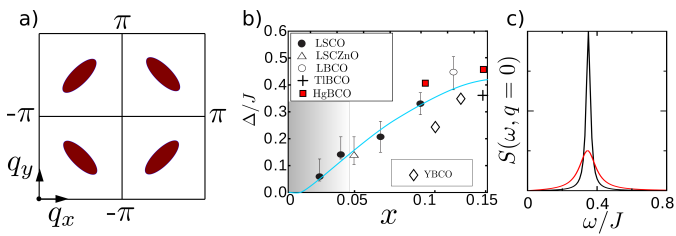}
  \caption{
    Panel (a): Small Fermi surface and momenta $q_x$, $q_y$ for magnetic
    scattering.
    Panel (b): The spin liquid gap versus doping
    (cyan line)~\cite{Kharkov2018a}.
    The theoretical curve corresponds to $\lambda=1$.
    Points show  experimental data. In different compounds values
    of $\lambda$ can be somewhat different and this explains scattering
    of experimental points.
    Panel (c): The $\omega$-scan of the spectral function at
    $q=0$. The black solid line illustrates the perfect $\delta$-function,
    and the dashed red line shows the temperature (disorder) broadened line.
  }
\label{fig:PD}
\end{figure}

We introduce superconductivity  in the theory \textit{ad hoc} via the
phenomenological d-wave SC gap
\begin{equation}
\label{DeltaSC2}
\Delta_{\bm k}=\Delta_{SC}\frac{1}{2}(\cos k_x - \cos k_y) \ .
\end{equation}
For details see Appendix C.
In our theory the $(\pi,\pi)$ resonance (the neck of the HG) is
unrelated to SC.
The resonance is a manifestation of the SL gap $\Delta$
which is practically the same in the N state and in the SC state.
Quite often in neutron scattering papers the energy of the neck of HG
is denoted by $E_{cross}$, this is the same as the SL gap, $\Delta=E_{cross}$.
The gap $\Delta$ was calculated in
Ref.~\cite{Kharkov2018a}
and is plotted versus doping in Fig.\ref{fig:PD}b.
The scattering of experimental points in this plot
is one of the manifestations of the observation {\bf ({O6})} which is explained
by proximity to the magnetic criticality.

We describe magnetic response at energy $\omega \lesssim J$
in terms of fluctuating staggered magnetisation ${\vec n}$.
The Green's function of the $\vec n$-field in the SL phase  reads, see
Appendix A3,
\begin{eqnarray}
D(\omega,\bm q)= 
  \ \frac{16J\sqrt{1-\mu x}}{\omega^2 - c^2q^2 - \Delta^2 - \Pi(\omega, \bm q) + i 0 }.
\label{eq:D}
\end{eqnarray}
Here 
$\Delta$ is the SL gap and $\Pi(\omega,q)$ is the magnon
polarisation operator.
The magnon speed is $c\approx c_0\sqrt{1-\mu x}$,
where $c_0\approx \sqrt{2}J$ is the magnon speed in the undoped compound.
The coefficient $\mu\approx 4$ has been calculated numerically within
the $t-J$ model~\cite{Kharkov2018a}.
The uncertainty of this calculation is about  $4 < \mu < 5$.
  The coefficient $\mu$ determines the dispersion slope softening at high
  energy. The softening is known  experimentally and therefore the value of
  $\mu$ can be also extracted from experimental data. This gives the same
  uncertainty interval $4 < \mu < 5$.

The magnetic spectral function 
is $S(\omega, \bm q)= - \frac{A}{16\pi J} {\mbox Im} D$.
The proportionality coefficient $A$ depends on the on-site magnetic moment,
Cu atomic form factor, etc. This coefficient is practically the same
for all hole doped cuprates, at least for the single layer cuprates.
The spectral function in the parent compound (N\'eel state, $x=0$) is
\begin{eqnarray}
  \label{SN}
  S_{Neel}(\omega, \bm q)= A \delta (\omega^2-c_0^2q^2).
\end{eqnarray}  
The holon polarisation operator $\Pi(\omega, \bm q)$ calculated in Appendix C
is a complex function of $\omega$ and  $\bm q$, however $\Pi(\omega,0)=0$.
Therefore, the spectral function at finite doping at $q=0$,
$S(\omega,0)= A \sqrt{1-\mu x}\delta (\omega^2-\Delta^2)$,
is a perfect $(\pi,\pi)$-resonance shown in Fig.\ref{fig:PD}c.
Impurities and finite temperature give rise to a broadening
illustrated by the red dashed line in Fig.\ref{fig:PD}b. Importantly,
the spectral weight is independent of the broadening and is
 defined by the SL gap $\Delta$ in the particular compound:
\begin{eqnarray}
  \label{SW}
  W=\int_0^{\infty}S(\omega,0)d\omega \approx  
  \frac{A \sqrt{1-\mu x}}{2\Delta}\ .
\end{eqnarray}

\section{Calculated q-scans of the spectral function at optimal doping}
\begin{figure}[h!]
  \includegraphics[scale=0.19]{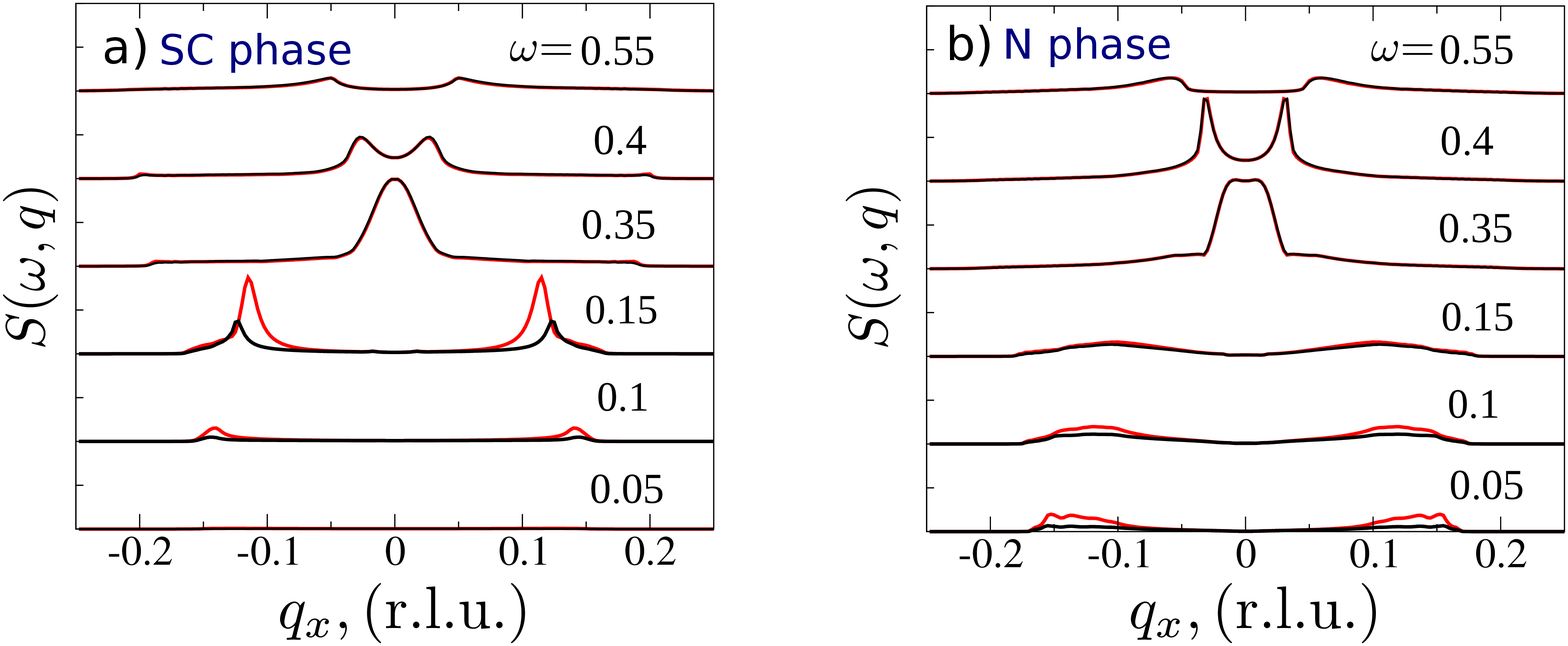}
  \caption{Calculated momentum scans of the spectral function
    for $x=0.16$, $\Delta=0.35J$, $\Delta_{SC}=0.2J$  and six different
    energies. Energies
in units of $J$
    are shown  near the respective lines.
    Panel (a) corresponds to the SC-state and
    panel (b) to the N-state. 
 Black curves correspond to $\lambda=1$ and red curves correspond to
$\lambda=1.1$. Black and red practically coincide for higher energies.
  }
\label{fig:spectr}
\end{figure}

\begin{figure}
  \includegraphics[scale=0.185]{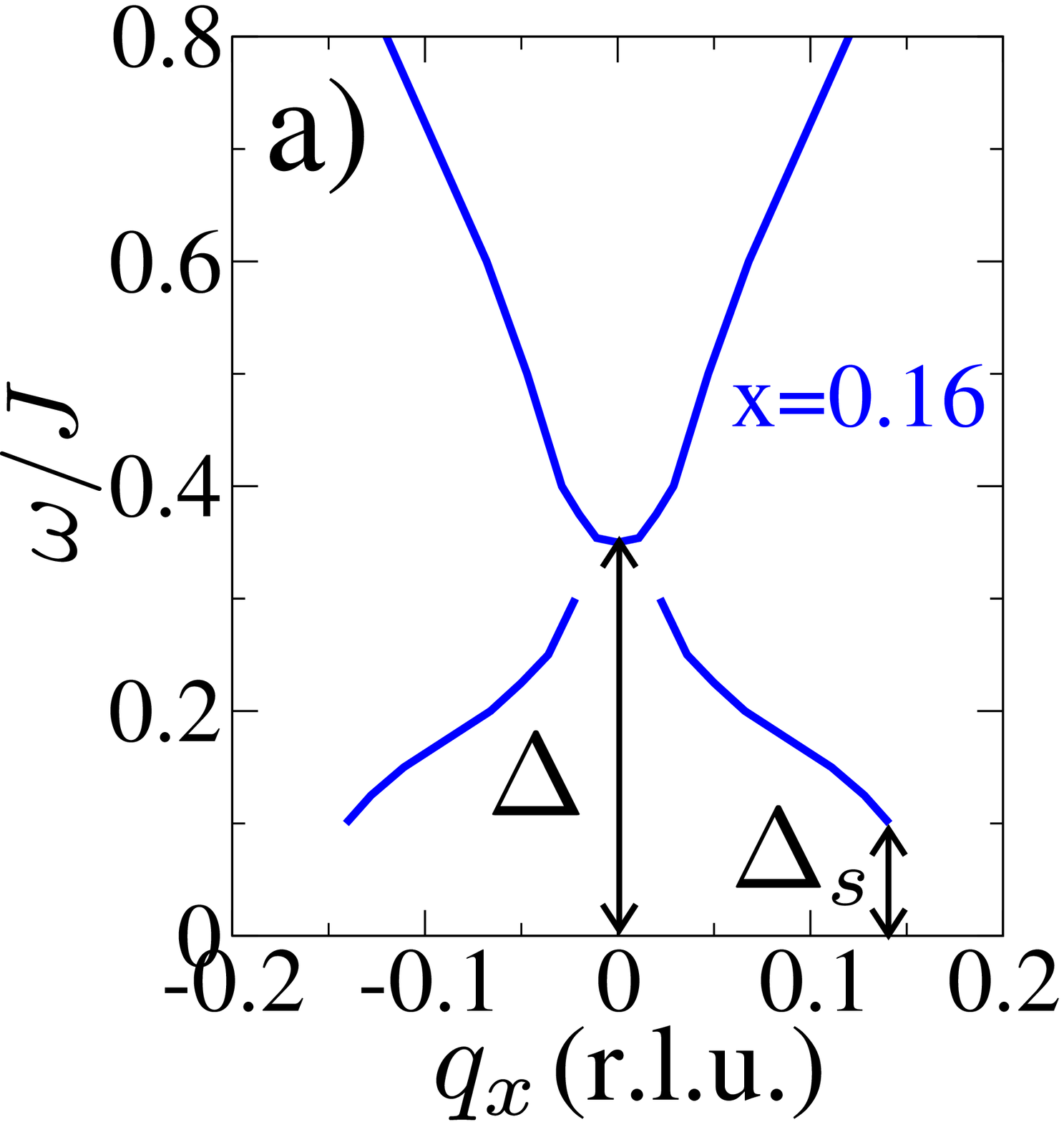}
  \includegraphics[scale=0.73]{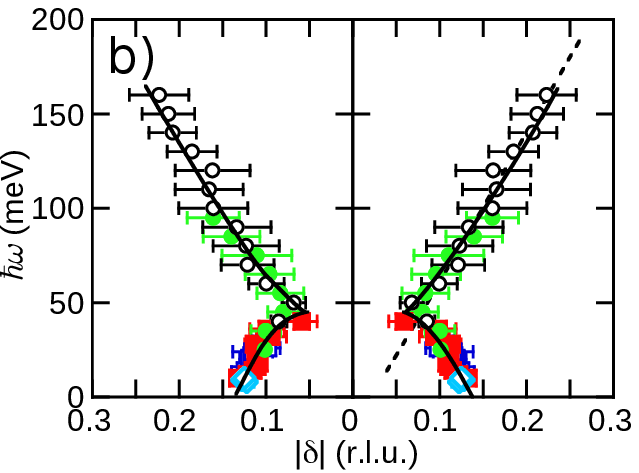}
\caption{Panel a: The HG dispersion for x=0.16 deduced from spectra in
Fig.\ref{fig:spectr}a. In this plot we indicate the
SL gap $\Delta$ and the spin pseudogap $\Delta_s$.
Panel b: The HG dispersion in La$_{1.84}$Sr$_{0.16}$CuO$_4$ from
Ref.~\cite{Vignolle07}
}
\label{HG0.16}
\end{figure}

In order to explain mechanism of the HG we consider separately the upper and
the lower part of the HG, see Fig.\ref{HGf}a. In a crude approximation
one can neglect the polarisation operator in
Eq.(\ref{eq:D}) for the upper part of the HG and this results in the dispersion
\begin{eqnarray}
  \label{uhg}
  \omega_q\approx \sqrt{\Delta^2+c_0^2(1-\mu x)q^2} \ .
\end{eqnarray}
Of course we can do better than this crude approximation.
In Fig.\ref{fig:spectr} we plot  $q_x$-scans of the spectral function
calculated numerically (for definition of axes see Fig.\ref{fig:PD}a).
The calculation accounts for the polarisation operator and is performed
for $x=0.16$ and six values of $\omega$.
The value of the SL gap
is taken from Fig.\ref{fig:PD}b, $\Delta=0.35J$,
and $\Delta_{SC}=0.2J \approx 25$meV,   see Appendix C.
 Panel (a) in Fig. \ref{fig:spectr} corresponds to the SC
state and panel (b) to the N state~\cite{com3}.
The plot demonstrates that the upper part of the HG,
$\omega > \Delta$, is only weakly sensitive to SC in agreement with
observation {\bf (O4)}.
In both SC and N phases the polarisation operator gives a broadening and
some asymmetry of peaks in the spectral function, but overall the upper part of HG is consistent with
crude Eq.(\ref{uhg}).
On the other hand  according to Fig.\ref{fig:spectr} sharp peaks in
the lower part of HG, $\omega < \Delta$,
exist in the SC state and disappear in the N state in agreement with
observation {\bf (O2)}.
However, the peaks are still present in the N-state,
they just become very broad due to the decay to the particle-hole continuum.
The spectra in Fig.\ref{fig:spectr}a correspond to the HG dispersion
plotted in Fig.\ref{HGf}a by the blue line. We plot the dispersion again
in Fig.\ref{HG0.16}a with indication of   SL gap $\Delta$ and the spin
pseudogap $\Delta_s$.
According to Fig.\ref{fig:spectr}a in the SC state the magnetic response is
strongly suppressed at low frequencies, $\omega \lesssim 0.1J$.
The suppression can be described by the spin pseudogap $\Delta_s$
indicated in Fig.\ref{HG0.16}a.  Unlike the true gap $\Delta$, the 
$\Delta_s$ is a pseudogap since at the incommensurate q-points there is
some response down to zero energy. 
In the N state, Fig.\ref{fig:spectr}b, the low energy response is strongly
enhanced.
The presence of the zero-frequency magnetic response in the vicinity
  of the  incommensurate wave vector $Q\approx 0.14$r.l.u.  results in a nonzero NMR relaxation
  rate.
  We illustrate  sensitivity of the spin pseudogap $\Delta_s$
  to the magnetic criticality parameter $\lambda$ by plotting in
  Fig.\ref{fig:spectr} the spectral function 
for two values of $\lambda$: $\lambda=1$ (black) and
$\lambda=1.1$ (red). These  values correspond to red and black squares on the phase diagram 
Fig.\ref{HGf}b.
Both in the SC and the N states the upper part of HG is not sensitive to
the small variation of $\lambda$.
Conversely, the lower part in the SC state, $\omega < \Delta$, is very
sensitive.
Naturally the spin pseudogap $\Delta_s$ is smaller at $\lambda=1.1$
compared to that at $\lambda=1$, since the former is closer to the
phase boundary in Fig.\ref{HGf}b.

\begin{figure}
\includegraphics[scale=0.18]{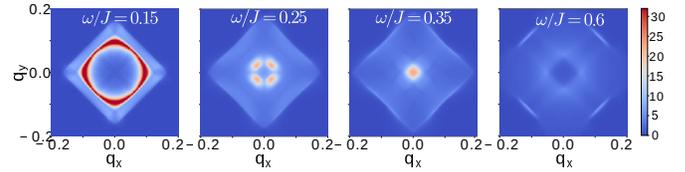}
\caption{Colour maps of fixed energy $(q_x,q_y)$-scans of  magnetic spectral
  function $S(\omega,q)$ for $x=0.16$ in SC phase.
  The magnetic criticality parameter $\lambda=1.1$,
  the SL gap is  $\Delta =0.35 J$, the SC gap is $\Delta_{SC}=0.2J$.
}
\label{map}
\end{figure}

Two dimensional ($q_x,q_y$) colour maps of the calculated structure factor
corresponding to the ``red'' spectra in Fig.\ref{fig:spectr}a
are presented in Fig.\ref{map}.
 The maps are close to  data
for La$_{1.84}$Sr$_{0.16}$CuO$_4$, Fig. 1 in Ref.~\cite{Vignolle07}.
Detail comparison shows that La$_{1.84}$Sr$_{0.16}$CuO$_4$ is slightly
more critical. Increasing $\lambda=1.1$ to  $\lambda \approx 1.15-1.2$
we can even better reproduce data of Ref.~\cite{Vignolle07}
However, we do not perform the fit due to the  reason
explained in the Section  VIII.

\section{Magnetic criticality mechanism of the hour glass dispersion}
\begin{figure}[h!]
  \includegraphics[scale=0.17]{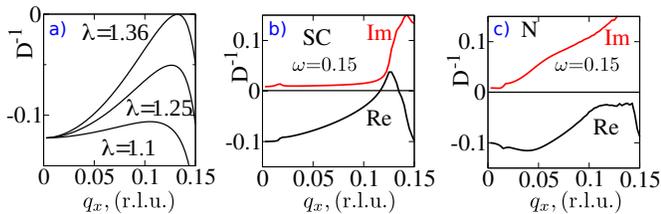}
  \caption{Denominators of the  Green's function (\ref{eq:D})
versus $q_x$ for $x=0.16$, $\Delta=0.35J$, $\Delta_{SC}=0.2J$.
Panel (a): SC state, $\omega=0$, three different values of $\lambda$.
Panel (b): SC state, $\omega=0.15J$, $\lambda=1.1$,
real (black) and imaginary (red) parts.
Panel (c): N state, $\omega=0.15J$, $\lambda=1.1$,
real (black) and imaginary (red) parts.
  }
\label{fig:denom}
\end{figure}

The results presented in the previous section indicate that the HG is driven
by the SL
Lifshitz magnetic criticality, SC plays a secondary role and leads to
the spectral line narrowing in the lower part of HG.
In order to elucidate this point  we plot in  Fig.\ref{fig:denom}
denominators of the Green's function (\ref{eq:D})
versus $q_x$ for $x=0.16$.
At $\omega=0$ the denominator is real and Fig.\ref{fig:denom}a displays 
the $\omega=0$ denominator in the SC state for three different values of
$\lambda$.
For $\lambda=1.1$ and $\lambda=1.25$ the denominator is negative indicating
stability of the SL phase. However, at $\lambda=1.36$ the denominator vanishes, $D^{-1}=0$, at $q_x\approx 0.135$ r.l.u. indicating condensation of the
spin spiral with this wave vector, see the phase diagram Fig.\ref{HGf}b.
In Fig.\ref{fig:denom}b we plot the denominator in the SC state for
$\omega=0.15J$ and $\lambda=1.1$,
and in Fig.\ref{fig:denom}c we plot the same denominator but in the N state.
In the SC state the Green's function denominator is close to zero at
$q_x\approx 0.12$ r.l.u. and this results in a narrow peak  in the
$\omega=0.15J$ red curve in Fig.\ref{fig:spectr}a.
In the N state the real part of the denominator is also small,
but the imaginary part is large, thus the magnetic critical
enhancement just results in a very broad structure in  the
$\omega=0.15J$ red curve in Fig.\ref{fig:spectr}b.

Thus the HG is a collective excitation  driven by the Lifshitz magnetic
criticality. The role of SC is just to suppress the decay rate
of the collective excitation.

\section{Is there a hole in the hour glass?}
Sometimes experimental HG dispersion is plotted with a hollow ``neck''
as it is shown in Fig.\ref{HG0.16}b.
We think that the hollow neck does not exist, but we understand how
it can mistakenly arise in the analysis of experimental data.
Fig.\ref{fig:spectr}a demonstrates pairs of narrow peaks for scans
above and below the HG neck and a broad peak at the neck $\omega=\Delta=0.35 J$.
An assumption that the broad peak consists of two narrow peaks
leads to the hollow neck. However, we believe this is wrong, 
the neck of the HG is intrinsically broad.
There is no  hole in HG.

\section{Calculated q-scans of the spectral function in the underdoped case.
Emergent incommensurate magnetic order}
Next we look at the lower doping, $x=0.1$. The SL gap according to
Fig.\ref{fig:PD}b is $\Delta=0.3J$.
We keep the value of the magnetic criticality parameter unchanged,
$\lambda=1.1$, but the value of the SC gap is reduced,
$\Delta_{SC}=0.1J \approx 13$meV,   see Appendix C.
Fig.\ref{fig:spectrA} presents calculated $q_x$-scans of the spectral
function for  six values of $\omega$. 
\begin{figure}[h!]
  \includegraphics[scale=0.19]{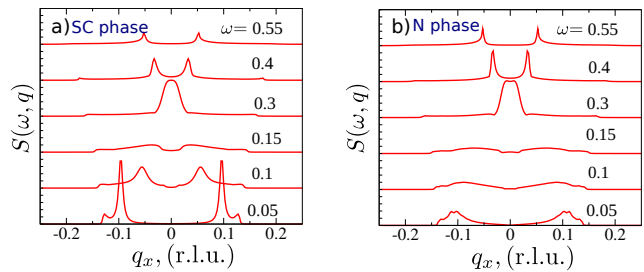}
  \caption{Calculated momentum scans of the spectral function
    for $x=0.1$, $\Delta=0.3J$, $\Delta_{SC}=0.1J$, $\lambda=1.1$,
    and six different energies.
    Energies in units of $J$
    are shown  near the respective lines.
    Panel (a) corresponds to the SC-state and
    panel (b) to the N-state.
  }
\label{fig:spectrA}
\end{figure}
The red circle on the phase diagram  Fig.\ref{HGf}b corresponding to these
 parameters  is located close to the critical line.
Therefore, the low frequency response in Fig.\ref{fig:spectrA} is strongly
enhanced compared to that in Fig.\ref{fig:spectr}.
In this case the spin pseudogap is practically zero, $\Delta_s\approx 0$.
Moreover, the lower part of HG becomes evident even in the N state
in agreement with the observation {\bf (O3)}.
Further decreasing of
doping would  result in an emergent static incommensurate magnetic
order in agreement with the observation {\bf (O5)}.

The HG dispersion deduced from Fig.\ref{fig:spectr}a and Fig.\ref{fig:spectrA}a, are plotted
in Fig.\ref{HGf}a by the blue and red lines, respectively.

\section{Wine Glass dispersion}
\begin{figure}[h!]
  \includegraphics[scale=0.2]{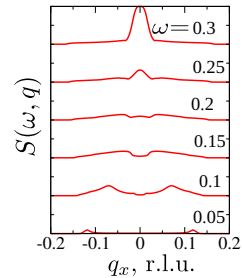} 
  \caption{
    Calculated momentum scans of the spectral function in the SC state
    for $x=0.1$, $\Delta=0.3J$, $\lambda=0.9$,  and 6 different energies.
    Energies in units of J
    are shown  near the respective lines.
}
\label{fig:WG}
\end{figure}

It is clear from the above discussion that decreasing of the magnetic
criticality parameter $\lambda$ reduces the intensity of the lower part of
the HG.
In Fig.\ref{fig:WG}  we present momentum scans for $\lambda=0.9$.
All other parameters are the same as that in Fig.\ref{fig:spectrA},
$x=0.1$, $\Delta=0.3J$, $\Delta_{SC}=0.1J$.
The blue star on the phase diagram  Fig.\ref{HGf}b corresponds to this
set of parameters.
In Fig.~\ref{fig:WG} the intensity in the lower part of the spectrum
is dramatically reduced compared to that in Fig.\ref{fig:spectrA}.
At the same time the upper part of the spectrum is practically the same.
Thus, reducing $\lambda$ one drives the HG dispersion
to the ``wine-glass'' regime  reported in Refs.~\cite{Chan2016,Chan16a}

 \section{Lifshitz spin liquid fingerprint relation}
 In the previous sections we have explained all the major HG
 observations {\bf (O1)-(O6)}
 listed in the introduction. Is there a further experimental confirmation
 of the developed theory? The answer is yes. The central point  of the theory
 is that the Lifshitz SL is very similar to the parent antiferromagnet.
 Most explicitly this point is reflected in Eqs.(\ref{SN}),(\ref{SW}).
 The  spectral weight W in the SL phase, Eq.(\ref{SW}), is expressed
via the coefficient A known from the parent antiferromagnet, Eq.(\ref{SN}).
 For this reason we call Eq.(\ref{SW}) the Lifshitz SL
 ``fingerprint'' relation. Let us compare this relation with
 experimental data.

\begin{figure}[h!]
  \includegraphics[scale=0.26]{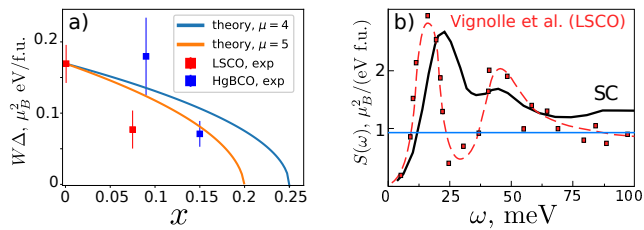}
  \caption{Panel (a): The SL  sum rule $ W \Delta =0.5A\sqrt{1-\mu x}$,
   Eq.(\ref{SW}),    versus doping.
    Two theoretical curves indicate uncertainty of $\mu$.
    The red points are extracted from La$_{2-x}$Sr$_{x}$CuO$_4$
    data~\cite{Matsuura2017} and the blue points from
    HgBa$_2$CuO$_{4+\delta}$ data~\cite{Chan2016, Chan16a}.
    Panel (b): Black line represents theoretical
     $q$-integrated spectral function $S(\omega)$ for 
    $x=0.16$, $\Delta=0.35J$, $\Delta_{SC}=0.2J$, $\lambda=1.1$.
     The red line is experimental $S(\omega)$ for 
     La$_{1.84}$Sr$_{0.16}$CuO$_4$, Ref.~\cite{Vignolle07}
    The blue horizontal line  shows $S(\omega)$ in  La$_2$CuO$_4$.
  }
\label{So}
\end{figure}

 Using Eq.(\ref{SN}) and
fitting the 5K data in Fig.4a of Ref.~\cite{Matsuura2017}  for undoped
La$_2$CuO$_4$ we find $A\approx 0.35\mu_B^2 eV/f.u.$
Hence, using Eq.(\ref{SW}) we plot in Fig.\ref{So}a theoretical
curves for the product $W \times \Delta$ versus doping. We know 
the value of
the coefficient $\mu$ in (\ref{SW}) only approximately, therefore we present curves for $\mu=4,5$ to indicate theoretical uncertainty.
At the same Fig.\ref{So}a  we present experimental points extracted from data
for La$_{2-x}$Sr$_{x}$CuO$_4$ (red)~\cite{Matsuura2017} and 
HgBa$_2$CuO$_{4+\delta}$ (blue)~\cite{Chan2016, Chan16a}. The red point at $x=0$
gives the normalisation of the theoretical curve. The agreement is
quite good, the data are consistent with the SL ``fingerprint'' relation.

It would be very interesting to perform a similar analysis for YBCO.
The compound has theoretical complications related to the double
layer structure
and to the oxygen chains, but these issues are probably resolvable.
The major problem is that there is not enough data
with  absolute normalisation of intensity.

The next point we address is the $q$-integrated spectral function
$S(\omega)=\int S(\omega,{\bm q})\frac{d^2q}{(2\pi)^2}$.
To be specific we take the same set of parameters as that in
Fig.\ref{fig:spectr}, $x=0.16$, $\Delta=0.35J$, $\lambda=1.1$,
$\Delta_{SC}=0.2J$.
The calculated  spectral function in the SC state is shown in
Fig.\ref{So}b by the black line. For normalisation we use the value of $A$
extracted from undoped La$_{2}$CuO$_4$ as described in the second  paragraph
of this section.
In the same Fig.\ref{So}b we plot the experimental curve (red) for
 La$_{1.84}$Sr$_{0.16}$CuO$_4$, Ref.\cite{Vignolle07}
 The  agreement between the theory and the experiment both in shape and
 in the absolute normalisation is good.
 The characteristic double hump structure has been also observed in
 YBa$_2$Cu$_3$O$_{6.5}$, Ref.\cite{Keimer1997a}
 As we pointed out above, a slight increase of criticality parameter,
 $\lambda=1.1 \to \lambda\approx 1.15-1.2$  
 would shift the low peak down close to the experimental position.
 In principle one can try to fit the data by changing $\lambda$.
 However, there are recent evidences~\cite{wagman2015,Xu2014}
indicating a phonon with energy $\omega \approx 18-20$meV.
The phonon  adds some intensity to the lower peak in Fig.~\ref{So}b.
Of course the phonon is not described by our theory. This is why we do
not fit data of Ref.\cite{Vignolle07} presented in Fig.~\ref{So}b.

 According to Eq.(\ref{SN}) the $q$-integrated spectral function
 in the parent antiferromagnet is $S(\omega)= \frac{A}{8\pi J^2}
 \approx 0.95\mu_B^2/(eV \ f.u.)$. This value is shown in 
 Fig.\ref{So}b by the blue horizontal line. In the energy interval
 above the neck of HG, $\omega=50-80$meV, this value coincides
 with $S(\omega)$ for La$_{1.84}$Sr$_{0.16}$CuO$_4$ presented in the
 same figure. This proves the point {\bf (F4)} listed
   in the Introduction.
   The same point is true for YBCO. Relevant data are presented
   in Fig.2a of Ref.\cite{Keimer1997a} Solid lines in this
   figure represent $S(\omega)$ in YBa$_2$Cu$_3$O$_{6.5}$ for three
   different temperature and the horizontal dashed line shows
   $S(\omega)$ in the parent antiferromagnet~\cite{Bourges2019}.
   From these data we conclude that in the interval $\omega=50-80$meV
   the structure factor $S(\omega)$ in YBa$_2$Cu$_3$O$_{6.5}$ and in the
   parent antiferromagnet has the same value.\\

\section{Conclusions}
We show that the hour-glass magnetic dispersion in underdoped cuprates
is driven by properties of the  Lifshitz magnetic critical spin liquid.
Superconductivity plays a secondary role and only responsible for the narrowing of the spectral lines.
We list the six major observations related to the hour-glass dispersion
and explain all of them.
We propose a spin liquid ``fingerprint relation'' and demonstrate that neutron scattering data support the relation.

\begin{acknowledgments}
{\it Acknowledgments}.
We thank G. Khaliullin for stimulating discussions.
We also thank 
P. Bourges, M. Fujita, M. Greven, S. M. Hayden, B. Keimer,
J. M. Tranquada, and C. Ulrich
for discussions and important communications.
The work has been supported by Australian Research Council No DP160103630.    
\end{acknowledgments}

\newpage

\appendix
\section{Effective action}
\subsection{Extended $t-J$ model}
The Hamiltonian of the extended $t-J$ model reads \cite{Anderson87, Emery87, Zhang88}
\begin{eqnarray}\label{eq:Ham_tJ}
H = -t\sum_{\langle ij \rangle} c^\dag_{i, \sigma} c_{j, \sigma} -t'\sum_{\langle \langle ij \rangle\rangle} c^\dag_{i, \sigma} c_{j, \sigma} - \nonumber \\
t''\sum_{\langle \langle \langle ij \rangle\rangle\rangle} c^\dag_{i, \sigma} c_{j, \sigma} + J \sum_{\langle i,j \rangle} \left[ \bm S_i\cdot \bm S_j - \frac{1}{4}N_i N_j \right],
\end{eqnarray}
where $c^\dag_{i\sigma}$ ($c_{i\sigma}$) is the creation (annihilation) operator for an electron with spin $\sigma=\uparrow,\downarrow$ at Cu site $i$; the operator of electron spin reads $\bm S_i = \frac{1}{2} c^\dag_{i\alpha} \bm\sigma_{\alpha\beta} c_{i\beta}$. The electron number density operator is $N_i = \sum_\sigma c^\dag_{i\sigma} c_{i\sigma}$, where $x$ is the hole doping, so that the sum rule $\langle N_i \rangle = 1-x$ is obeyed.
In addition to Hamiltonian (\ref{eq:Ham_tJ}) there is the no double occupancy constraint,
which accounts for a strong electron-electron on-site repulsion. 
The value of superexchange is approximately the same for all cuprates,
$J \approx 125$meV. The superexchange has been directly measured and
shown to be independent of doping~\cite{Imai1993}.
While in Eq.(\ref{eq:Ham_tJ}) we present only three hopping matrix elements,
the nearest site hopping $t$,  the next nearest site hopping $t'$, 
and the next next nearest site hopping $t''$, we know from
LDA calculations~\cite{andersen95} that more distant hoppings, $t^{(3)}$,
$t^{(4)}$ and even  $t^{(5)}$ are also significant. 
Unfortunately
values of the hopping matrix elements cannot be directly measured.
It is widely believed that the value $ t \approx 400meV \approx 3J $ is
reliable and common for all cuprates, we use this value in the present work. 
However, values of the distant hopping matrix elements are rather uncertain
and can vary from one family to another.

The Fermi surface of a lightly doped extended $t-J$ model consists 
of Fermi pockets shown in Fig.\ref{fig:PD}a and centred
at the nodal points $\mathbf{k}_{0}=(\pm \pi /2,\pm \pi /2)$, and $\mathbf{k}_{0}=(\pm \pi /2,\mp \pi /2)$.
The single hole dispersion can be parameterised as~\cite{Sushkov1997}
\begin{eqnarray}
\label{disp}
&&\epsilon_{\bm k}=\beta_1(\gamma_{\bm k}^+)^2+\beta_2(\gamma_{\bm k}^-)^2  , \ \ 
\gamma_{\bm k}^{\pm}=\frac{1}{2}(\cos k_x \pm \cos k_y),\nonumber\\
&&\epsilon_{\bm k}\approx \beta_1\frac{p_1^2}{2}+\beta_2\frac{p_2^2}{2}.
\end{eqnarray}
Here  ${\bf p}={\bf k}-\mathbf{k}_{0}$, see Fig.\ref{fig:PD}a.
We set the lattice spacing equal to unity,
$a= 3.81 \AA \rightarrow 1 $.
The second line in Eq.(\ref{disp}) corresponds to the quadratic expansion of
the fermion dispersion in the vicinity of the centers of Fermi pockets. 
The ellipticity of the holon pocket is $\sqrt{\beta_1/\beta_2}$.
The Fermi energy is related to doping as
\begin{eqnarray}
\label{eF}
&&\epsilon_F\approx \pi\beta x, \nonumber\\
&&\beta=\sqrt{\beta_1\beta_2}=\frac{1}{m^*}.
\end{eqnarray}
Values of the inverse effective masses $\beta_1$, $\beta_2$
follow from Hamiltonian (\ref{eq:Ham_tJ}). They have been
calculated using self-consistent Born approximation (SCBA),
see Refs.\cite{Sushkov1997,Sushkov04}
The values strongly depend on distant hopping parameters which are
essentially unknown, even $t^{(3)}-t^{(5)}$ significantly influence SCBA results.
For illustration we present here values of $\beta_1$, $\beta_2$ obtained for
several sets of the distant hopping parameters.
We consider only the sets that result in positive $\beta_1$ and $\beta_2$.
For the ``pure'' $t-J$ model,  $t'=t''=t^{(3)}=t^{(4)}=t^{(5)}=0$, the
values are $\beta_1=1.96J$, $\beta_2=0.30J$. 
For the set $t'\approx0.23J$, $t''=0$, $t^{(3)}=t^{(4)}=t^{(5)}=0$
one gets the Van Hove singularity, $\beta_2=0$.
On the other hand in the limit $t''\gg t',J$ the
inverse masses are very large $\beta_1\approx \beta_2 \approx 8t''\gg J$.
For the middle of the LDA range, Ref.\cite{andersen95},
$t'= -0.5J $, $t''= 0.5J $, $t^{(3)}=t^{(4)}=t^{(5)}=0$
the inverse masses are $\beta_1\approx 2.76J$, $\beta_2 \approx 2.62J$.
While we can claim that $\beta=\sqrt{\beta_1\beta_2}\sim 2$,
the so strong dependence on unknown parameters indicates that in the end
the effective masses and especially the ellipticity of the Fermi pocket
have to be taken from experiment.

Even more important than the effective masses is the dimensionless magnetic
criticality parameter~\cite{Sushkov04,Milstein2008}
\begin{equation}
\label{ll}
\lambda = \frac{8 g^2}{\pi \sqrt{\beta_1 \beta_2}}.
\end{equation}
Here $g=Zt$ is the magnon-holon coupling constant, $Z$ is the holon
quasiparticle residue. In theory by varying $t'$ and $t''$ one can vary
$\lambda$ from zero to infinity. In the large $t''$ limit,
$t''\gg t',J$, the parameter is very small, $\lambda \to 0$.
On the other hand near the Van Hove singularity, $\beta_2\to 0$, the parameter
is very large $\lambda \to \infty$. 
For the ``pure'' $t-J$ model,  $t'=t''=t^{(3)}=t^{(4)}=t^{(5)}=0$
the criticality parameter value is $\lambda=2.51$.
For the middle of the LDA range~\cite{andersen95},
$t'= -0.5J $, $t''= 0.5J $, $t^{(3)}=t^{(4)}=t^{(5)}=0$,
the criticality parameter value is $\lambda=1.1$.
Within the overall LDA range of the hopping
parameters~\cite{andersen95} $\lambda$ varies from 1 to 2.
Numerically the difference between $\lambda=2$ and $\lambda=1$ is
not that large, but physically the difference is enormous.
The value $\lambda \geq 2$ implies that the system is unstable with respect
to the phase separation, see Ref.~\cite{Chubukov95} and  Ref.\cite{Sushkov04}
So, the ``pure'' $t-J$ model with $\lambda \approx 2.5$ is unstable
and hence inconsistent with experiment.
On the other hand the value $\lambda=1$ corresponds to the stable spin liquid
phase which is perfectly consistent with experiment, see Fig.1b
and Ref.~\cite{Kharkov2018a}
While from LDA+SCBA we can claim that $\lambda \sim 1$,
the strong dependence on unknown parameters indicates that in the end
 the value of $\lambda$ must be taken from experiment.
 Based on the phase diagram Fig.1b  we see that values
 $0.8 < \lambda < 1.3$ are generally consistent with data.\\

To summarise this section: we base our analysis on the extended $t-J$ model and use the value
$J\approx 125$ meV known from experiment. In the calculation we
use the value $t=3J$, we have checked that a variation of t within
$2.5J < t < 3.5J$ influences our results very weakly.
However, a  variation of distant hopping matrix elements, $t'$, $t''$,
$t^{(3)}$,...  has an enormous effect on physics. Variation of these
matrix elements within the window given by LDA calculations~\cite{andersen95}
can drive the system from the Neel state through the spin liquid state
to the spin spiral state and even to the phase separation.
Based on the spin liquid theory we conclude
that the range $0.8 < \lambda < 1.3$ is consistent with experimental
observations, so in our calculations we use this range.
Specifically in the paper we present results for $\lambda=1.1$,
$\lambda=1$, and $\lambda=0.9$ 
to demonstrate sensitivity to the criticality parameter.
The value of the effective mass is less important, in the paper we present results for
$\beta=\sqrt{\beta_1\beta_2}=2J$ ($m^*=2.1m_e$) and $\beta_1/\beta_2=6$.
We have checked that the set $\beta=\sqrt{\beta_1\beta_2}=3J$ and
$\beta_1/\beta_2=4$ results in practically the same answers.

\subsection{Quantum field theory: the low energy limit of the extended
  $t-J$ model}\label{sec:LLL}
\begin{figure}[h!]
\includegraphics[width=0.3\textwidth]{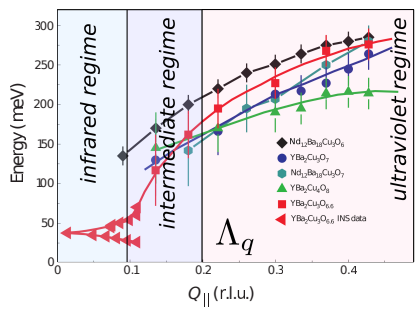}
\caption{Magnetic dispersion along the $(1,0)$ direction. Points show combined data on resonant 
inelastic X-ray scattering  and inelastic neutron scattering
in NdBCO and YBCO at $T=15$K, Ref. \cite{LeTacon11} 
Vertical lines separate three different regimes that we call  ``infrared regime'', ``intermediate regime'', and
``ultraviolet regime''.
}
\label{fig:RIXS_magnon}
\end{figure}

While the $t-J$ model is the low energy reduction of the three band
Hubbard model, the total energy
range in the $t-J$ model, $\Delta\epsilon \sim 8t \sim 24J\approx 3$eV,
is still very large.
Account for quantum fluctuations at lower energy scales practically
unavoidably requires a quantum field theory approach.
Theoretical arguments explaining this point have been discussed in
several theoretical papers including our recent work~\cite{Kharkov2018a}.
Here we repeat only experimental arguments supporting this statement.
In Fig.\ref{fig:RIXS_magnon} we present  magnetic dispersion along the
$(1,0)$ direction 
taken from Ref. \cite{LeTacon11} The dispersion is based on combined data on resonant 
inelastic X-ray scattering  and inelastic neutron scattering.
The data indicate three  distinct regimes separated in
Fig.\ref{fig:RIXS_magnon} by vertical lines.
In the ``ultraviolet regime'' the dispersion only very weakly depends on
doping, practically  independent. This is where our ``fact'' {(\bf F2})
in the Introductions comes from.
In the ``intermediate regime'' there is a significant softening with doping
and the most  dramatic doping dependence takes place in the ``infrared regime''.
The low energy effective field theory is relevant to the
``infrared'' and the ``intermediate'' regimes.
In these regimes energies of magnetic excitations and energies of holons are
small,
$\omega, \epsilon < 2J$. On the other hand in the ``ultraviolet regime''
the energies are large,
$\omega \sim 2J$ and $8t > \epsilon \gtrsim 2J$.
The field theory has the ultraviolet cutoff $\Lambda_q$ which is the upper
edge of the ``intermediate regime'' as it is indicated in
Fig.\ref{fig:RIXS_magnon}.
The value of the cutoff that follows from the data is
$\Lambda_q\sim 0.2$r.l.u). The same value follows from the theory,
see Ref.~\cite{Kharkov2018a}

The low energy Lagrangian of the $t-J$ model was first 
derived in Ref.\cite{Shraiman90} with some important terms missing.
The full effective Lagrangian was derived in Ref.\cite{Milstein2008}
This approach necessarily requires an introduction of two checkerboard
sublattices, independent of  whether there is a long range AFM order or
the order does not exist.
The two checkerboard sublattices allow us to avoid a double counting of
quantum states in the case
when spin and charge are separated.
A holon carries charge and does not carry  spin, but it can be located at
one of the
sublattices and this is described by the pseudospin $1/2$.
Due to the checkerboard sublattices the Brillouin zone coincides with
magnetic   
Brillouin zone (MBZ) even in the absence of a long range AFM order.
Therefore, there are four half-pockets in Fig.\ref{fig:PD}a or two full pockets
within MBZ.
Finally, the Lagrangian reads\cite{Milstein2008}
\begin{eqnarray} 
\label{eq:LL}
{\cal L}&=&\frac{\chi_{\perp}}{2}{\dot{\vec n}}^2- 
\frac{\rho_s}{2}\left({\bm \nabla}{\vec n}\right)^2\\
&+&\sum_{\alpha}\left\{ \frac{i}{2}
\left[\psi^{\dag}_{\alpha}{{\cal D}_t \psi}_{\alpha}-
{({\cal D}_t \psi_{\alpha})}^{\dag}\psi_{\alpha}\right] - \psi^{\dag}_{\alpha}\epsilon_{\alpha}({\bf \cal P})\psi_{\alpha} \right.\nonumber\\
&+&\left. \sqrt{2}g (\psi^{\dag}_{\alpha}{\vec \sigma}\psi_{\alpha})
\cdot\left[{\vec n} \times ({\bm e}_{\alpha}\cdot{\bm \nabla}){\vec n}\right]\right\} \  .
\nonumber
\end{eqnarray}
Fermions (holons) are described by a spinor $\psi_\alpha$ with the pseudospin
$1/2$, and the vector of staggered magnetization ${\bm n}$ normalised as
$\bm n^2=1$ corresponds to localised spins at Cu sites.
The first line in (\ref{eq:LL}) is $O(3)$ nonlinear sigma model that describes
spin dynamics, the second line is the Lagrangian for non-interacting holons. 
The long covariant derivatives in Eq. (\ref{eq:LL}) are defined as
\begin{eqnarray}
&\mathcal{ \bm P} = -i\bm \nabla + \frac{1}{2}\vec\sigma\cdot [\vec n \times \bm \nabla \vec n],\nonumber\\
&\mathcal{D}_t = \partial_t + \frac{1}{2} \vec\sigma \cdot [\vec n \times  \partial_t \vec {n}].\nonumber
\end{eqnarray}
The index $\alpha=1,2$ enumerates two full holon pockets in Fig.\ref{fig:PD}a.
The term in the bottom line in Eq. (\ref{eq:LL}) describes the coupling
between holons and the staggered magnetisation. 
Pauli matrices $\bm\sigma$ in Eq. (\ref{eq:LL}) act on the holon's pseudospin
and $\bm e_\alpha = 1/\sqrt{2}(1,\pm1)$ 
denotes a unit vector orthogonal to the face of the MBZ where the holon
is located. The coupling constant $g$ enters Eq.(\ref{ll}).

The Lagrangian (\ref{eq:LL}) is fully equivalent to the $t-J$ model
Hamiltonian (\ref{eq:Ham_tJ}). In essence Eq. (\ref{eq:LL}) originates from the Hamiltonian
(\ref{eq:Ham_tJ}) rewritten in notations convenient for analysis of the low energy physics.
The relation between (\ref{eq:LL}) and (\ref{eq:Ham_tJ}) is the same
as that between the nonlinear $\sigma$-model and the Heisenberg
antiferromagnetic model on square lattice.
The Lagrangian (\ref{eq:LL}) contains five parameters, $\chi_\perp$,  $\rho_s$,
$\beta_1$, $\beta_2$, and $g$. Of course they can be expressed in terms
of parameters of the ``parent'' $t-J$ model. We have already discussed
$\beta_1$ and $\beta_2$. The coupling $g=Zt$ is related to $\lambda$,
see Eq.(\ref{ll}), so it is also already discussed.
In the limit $x\to 0$ the $\sigma$-model
parameters $\chi_\perp$ and   $\rho_s$ coincide with that of the 2D Heisenberg model on the square lattice, 
$\chi_\perp=1/8J$, $\rho_s=J/4$, up to an overall scalar prefactor in Lagranian (\ref{eq:LL})  due to the renormalization of the spin magnitude by quantum fluctuations. The magnon speed is 
\begin{eqnarray}
\label{c0}
c_0=\sqrt{\rho_s/\chi_{\perp}}=\sqrt{2}J .
\end{eqnarray}

\subsection{Dependence of the Lagrangian parameters on doping}
The small parameter of our theory is doping, $x \ll 1$.
The Lagrangian parameters in subsection A2 are written in the
limit $x\to 0$. Here we discuss the $x$-dependence of the parameters up to
the linear in $x$ approximation. The first effect is renormalization of the
$\sigma$-model parameters due to fermionic fluctuations at the high energy
scale, $E\sim 8t \sim 24J$, see Ref.~\cite{Kharkov2018a}
\begin{eqnarray}
\label{ren}
&&\chi_\perp=\frac{1}{8J} \to \frac{1}{8J}\nonumber\\
&&\rho_s= \frac{J}{4}\to \frac{J}{4}(1-\mu x)\nonumber\\
&&c = \sqrt{\frac{\rho_s}{\chi_{\perp}}} \to c_0\sqrt{1-\mu x}
\end{eqnarray}
Note that the numerical coefficient $\mu\approx 4$ in the doping dependent prefactor is known only approximately, see Ref.\cite{Kharkov2018a}

The magnon Green's function generated
by (\ref{eq:LL}) above the spin liquid energy scale $\Delta$, $\Delta < \omega \lesssim J$ after taking into account Eqs.(\ref{ren}) reads
\begin{equation}
\label{gfa}
D(\omega, {\bm q})\propto \frac{\chi_{\perp}^{-1}}{\omega^2-c_0^2(1-\mu x) \bm q^2+i0} \ .
\end{equation}
Hence the sum rule for the spin structure factor is
\begin{eqnarray}
  \label{sr}
  \int S(\omega, {\bm q}) d\omega d^2q \propto \int Im D(\omega, {\bm q})
d\omega d^2q \propto \frac{\Lambda_q}{\sqrt{1-\mu x}}
\end{eqnarray}
The integration is performed in limits $0 < \omega < \infty$,
$0 < q < \Lambda_q$, where $\Lambda_q \approx 1.2\approx 2\pi*0.2(r.l.u.)$
 (we remind that we set the lattice spacing equal to unity)
 is the ultraviolet cutoff of the theory.
 Eq.(\ref{sr}) implies that the spin sum rule is increasing with doping.
 Obviously the sum rule should be a doping independent constant. This implies that magnetic fluctuations
at the scale $\omega \sim J \gg \Delta$ must generate the magnon
quasiparticle residue
\begin{eqnarray}
  Z_x=\sqrt{1-\mu x} \ .
\end{eqnarray}
  Eqs.(\ref{gfa}),(\ref{sr}) must be
multiplied by the residue and this makes the sum rule doping independent.

\section{The frustration mechanism behind the Lifshitz spin liquid}
Here we explain the mechanism of the Lifshitz SL without going to
technical details. The details are presented in Ref.\cite{Kharkov2018a}
The easiest way to understand how the Lifshitz SL arises due to frustration
of spins by
mobile holons is to stay in  the N\'eel phase, $\lambda <1$,  and 
to increase $\lambda$,  see the phase diagram Fig.\ref{HGf}b.
In the N\'eel phase there is a collinear long range order
$\langle n_z\rangle\ne 0$, and there is a quantum fluctuation
$\langle n_{\perp}^2\rangle=\langle n_{x}^2\rangle + \langle n_{y}^2\rangle$.
A calculation at small doping x gives the following answer for the fluctuation
\begin{eqnarray}
\label{np2}
&&\langle \vec n_{\perp}^2\rangle=\frac{x \beta}{2\rho_s}\ln\left(\frac{1}{1-\lambda}\right) + const \ .
\end{eqnarray}
When $\lambda$ is sufficiently close to unity the fluctuation is very large
and this results  in the quantum melting of the long range N\'eel order.
This explains the N\'eel - Lifshitz SL transition line in 
Fig.\ref{HGf}b. Similar arguments lead the Lifshitz SL - Spin Spiral
transition line on the phase diagram, see Ref.~\cite{Kharkov2018a}

\section{Effect of superconductivity}
With account of $Z_x$ the magnetic Green's function in the SL phase
reads~\cite{Kharkov2018a}
\begin{eqnarray}
  D(\omega,\bm q) &=& -i\int d^2 r dt e^{i\omega t + i qr}
  \langle T\left\{\bm n_i(\bm r, t) \cdot \bm n_i(0,0)\right\}\rangle
  \nonumber\\&=& \frac{2}{\chi_{\perp}}
  \ \frac{Z_x}{\omega^2 - c^2q^2 - \Delta^2 - \Pi(\omega, \bm q) + i 0 }
\label{eq:D1}
\end{eqnarray}
Here $\Delta$ is the SL gap and $\Pi(\omega,q)$ is the magnon
polarisation operator (fermionic loop).
The magnon-holon interaction  is given by the Lagrangian (\ref{eq:LL}).
Hence, a calculation of the magnon polarisation operator
is relatively straightforward.
In the calculation one can disregard superconducting pairing  of holons
or alternatively take the pairing into account.
Results without the pairing we call the ``normal state results''.

In the present work we introduce superconductivity in the theory \textit{ad hoc} via the phenomenological d-wave SC gap.
We use the simplest parametrization for the gap
\begin{equation}
\label{DeltaSC2}
  \Delta_{\bm k}=\Delta_{SC}\gamma_{\bm k}^- \ .
\end{equation}
In our numerical calculations we used the following values of the SC gap
\begin{eqnarray}
\label{DeltaSC3}
&& x=0.16: \ \ \ \Delta_{SC}=0.2J\nonumber\\
&& x=0.10: \ \ \ \Delta_{SC}=0.1J
\end{eqnarray}

Expressed in terms of parameters of the Lagrangian (\ref{eq:LL}) 
the zero temperature polarisation operator in the SC phase
reads~\cite{sushkov1996}
\begin{eqnarray}
\label{eq:P}
  &&  \Pi(\omega,q) =  \frac{2\pi\lambda c^2}{m^*} \sum_{\alpha=1,2} q_{\alpha}^2 \int \frac{d^2 k}{(2\pi)^2} \left\{ v^2_{\bm k} u^2_{\bm {k+q}}
\right.\\
&&\left.  +
  u_{\bm k} v_{\bm k} u_{\bm {k+q}} v_{\bm {k+q}}\right\}
  \left[\frac{1}{\omega - E_{\bm k} - E_{\bm {k+q}} + i 0} +
(\omega \to -\omega)\right], \nonumber
\end{eqnarray}
where $q_{\alpha}={\bm q}\cdot{\bm e}_{\alpha}$ and
$u_{\bm k}$ and $v_{\bm k}$ are Bogoliubov parameters:
\begin{eqnarray}
  &&u_{\bm k} = \sqrt{\frac{1}{2}\left( 1 + \frac{\xi_{\bm k}}{E_{\bm k}}\right)}
  \nonumber\\
&&v_{\bm k} = sign(\gamma_{\bm k}^-)\sqrt{\frac{1}{2}\left( 1 - \frac{\xi_{\bm k}}{E_{\bm k}}\right)}.
\end{eqnarray}
The  quasiparticle dispersion reads
\begin{equation}
  E_{\bm k} = \sqrt{\Delta_{\bm k}^2 + \xi_{\bm k}^2}, \quad
  \xi_{\bm k} = \epsilon_{\bm k} - \mu,
\end{equation}
where the chemical potential $\mu$  is defined by the condition
\begin{eqnarray}
  x = 2\sum_{\alpha=1,2} \int \frac{d^2k}{(2\pi)^2} v^2_{\bm k} \ .
\end{eqnarray}  
Numerical evaluation of the polarization operator
(\ref{eq:P}) is straightforward and we use it in the present work.


\begin{thebibliography}{99}
\bibitem{com3} In the SC state we set temperature $T=0$ and account for the
  SC gap. In the N-state we set the SC gap equal to zero, but still have T=0.
  This implies that our N state corresponds to the state with T just above
  $T_c$.
\end{thebibliography}
\end{document}